\documentclass{aa}          %normal two column style
\usepackage{graphicx} 

\def\LS{$L_{\odot}$}

\def\erg{erg cm$^{-2}$ s$^{-1}$}

\def\11{$\times$10$^{-11}$~erg cm$^{-2}$ s$^{-1}$} 
\def\cm2{cm$^{-2}$}
\def\22{$\times$10$^{22}$~cm$^{-2}$}
\def\23{$\times$10$^{23}$~cm$^{-2}$}

\def\III{\,{\sc iii}}

\begin{document} 
\title{ The high energy X--ray tail of $\eta$ Car revealed by 
BeppoSAX\thanks{
Based on space observations collected with the BeppoSAX X--Ray 
Astronomy Satellite which is a program of the Agenzia Spaziale
Italiana with participation of the Netherlands Agency for 
Aerospace Programs.
}}

\author{
R. F. Viotti 
\inst{1}, 
L. A. Antonelli \inst{2}, 
C. Rossi \inst{3}, 
S. Rebecchi \inst{4} 
}
\offprints{Roberto Viotti, e-mail: uvspace@rm.iasf.cnr.it} 
\institute{
Istituto di Astrofisica Spaziale e Fisica Cosmica, 
CNR, Area di Ricerca Tor Vergata,
Via del Fosso del Cavaliere 100, 00133 Roma, Italy 
\and 
INAF--Osservatorio Astronomico di Roma, 
Via di Frascati 33, 00040 Monte Porzio Catone (Roma), Italy 
\and
Dipartimento di Fisica, Universit\`a La Sapienza, 
Piazzale Aldo Moro 3, 00185 Roma, Italy 
\and
ASI Science Data Center (ASDC), c/o ESA-ESRIN, Via Galileo Galilei,
00044 Frascati (Roma), Italy 
}
\date{ Received  / Accepted }

\abstract{We report on the June 2000 long (100 ks) 
BeppoSAX exposure that has unveiled above 10 keV 
a new very high energy component of the X--ray spectrum 
of $\eta$ Car extending to at least 50 keV. 
We find that the 2--150 keV spectrum is best reproduced 
by a thermal $+$ non--thermal model.  
The thermal component dominates the 2--10 keV spectral 
range with kT$_h$=5.5$\pm$0.3 keV and log\,NH$_h$=22.68$\pm$0.01.
The spectrum displays a prominent iron emission line centred 
at 6.70 keV. Its equivalent width of 0.94 keV,
if produced by the thermal source, gives a slightly 
sub--solar iron abundance ([Fe/H]=$-$0.15$\pm$0.02). 
The high energy tail above 10 keV is best fitted by a power law 
with a photon index of 2.42$\pm$0.04. 
The integrated 13--150 keV luminosity of $\sim$12 \LS\ is 
comparable to that of the 2--10 keV thermal component (19 \LS).
The present result can be explained, 
in the $\eta$ Car binary star scenario,
by Comptonisation of low frequency radiation by high energy 
electrons, probably generated in the colliding wind shock front, 
or in instabilities in the wind of the S Dor primary star. 
It is possible that the high energy tail had
largely weakened near the minimum of the 5.53 yr cycle. 
With respect to the thermal component, 
it probably has a longer recovering time like 
that of the highest excitation optical emission lines. 
Both features can be associated with the large absorption 
measured by BeppoSAX at phase 0.05. 
}

\titlerunning{The high X-ray energy tail of $\eta$ Car}
\authorrunning{Viotti R.F. et al.}

\maketitle

\section{Introduction}
The peculiar southern object $\eta$ Car 
is one of the most remarkable variables in our Galaxy due to 
dramatic changes in its brightness.
In 1843 it was the second brightest star in the sky, 
then suffered a deep fading down to the 
eighth magnitude by the end of the 19th century 
(e.g. Viotti 1995). 
During the last century $\eta$ Car was slowly and irregularly 
re--brightening up to the present V$\simeq$5. 
Presently, according to the current distance estimates, $\eta$ Car  
has a bolometric magnitude around 5$\times$10$^6$ L$_{\odot}$ 
(2$\times$10$^{40}$ erg s$^{-1}$, e.g. Hillier et al. 2002). 
A mass loss rate of 10$^{-3/-4}$ M$_{\odot}$ yr$^{-1}$ 
or larger has been estimated from observations 
(e.g. Hillier et al. 2002, van Boekel et al. 2003, 
Pittard \& Corcoran 2002, Andriesse et al. 1978). 

Optical spectroscopic observations unveiled a peculiar 
cyclic behaviour, showing 
regularly repeating excitation minima, with a period 
of 5.53 years (Damineli et al. 2000). 
A similar behaviour was also found at other wavelength bands, 
from radio to X--rays, which is commonly interpreted in terms 
of a highly eccentric binary model composed of an S Dor--type 
very luminous primary star, and an unseen early type 
secondary star. 
The binary system interacts through colliding winds 
producing the observed $\eta$ Car's high temperature, 
luminous X--ray emission (e.g. Ishibashi et al. 1999, 
Corcoran et al. 2001). 

Recently, thanks to the BeppoSAX unique broad--band X--ray 
coverage we were able to detect, for the first time, $\eta$ Car 
above 10 keV (Viotti et al. 1998; Viotti et al. 2002a, Paper I). 
We reported the December 1996 observation showing 
a 13--20 keV flux in excess with respect to the extrapolated 
5 keV thermal spectrum that dominates the 2--10 keV range. 
The presence of a high energy tail was confirmed
by the following BeppoSAX observations of 31 December 
1999--2 January 2000 (Rebecchi et al. 2001). 
In particular, in June 2000 a 100 ks exposure unveiled 
that the tail was probably non--thermal and extending to at 
least 50 keV (Viotti et al. 2002b). 
In this work we analyse in detail the latter observation 
in order to investigate the origin of these very
high energy photons, and compare with the previous 
BeppoSAX observations of $\eta$ Car. The results are 
summarised in Table 1. 

%======================= Figure 1 =========================
\begin{figure*}
\centering 
\includegraphics[height=17cm,angle=270]{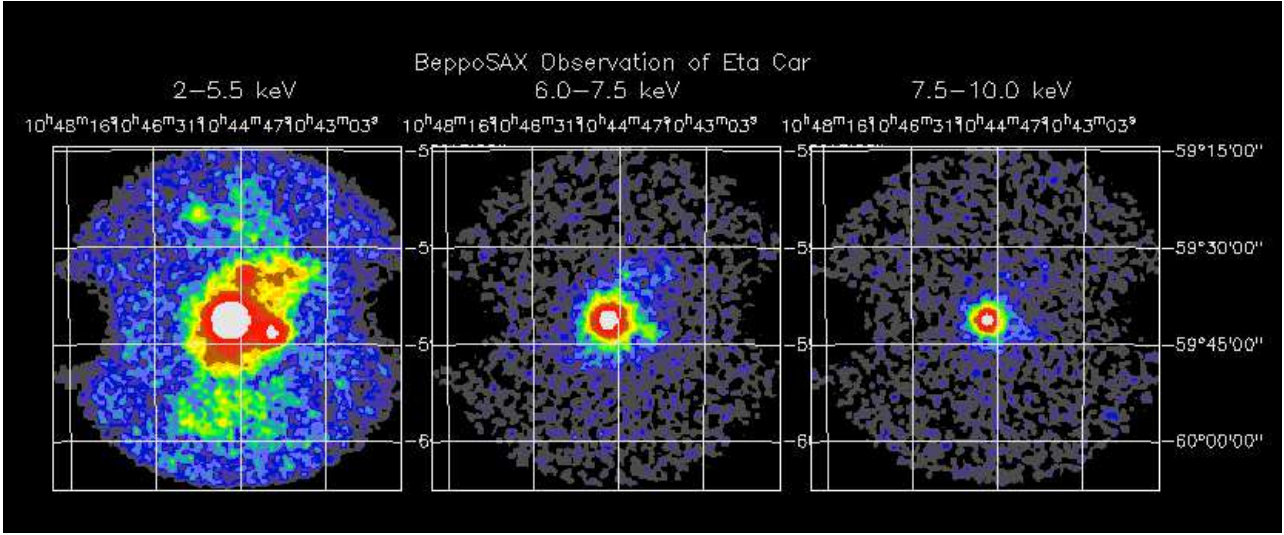} 
\caption{The BeppoSAX MECS images of the $\eta$ Car region on 
2000 June 21--23, in three energy bands: (from left to right) 
2--5.5 keV (softer), 6--7.5 keV (iron line), and 7.5--10 keV 
(harder). North is on the top and east to the left; 
the size of the image field is 54 arcmin. 
The second brightest source to the west of $\eta$ Car 
is the W--R star HD 93162 (WR25).}
\end{figure*}
%============================================================

\section{Observations}
The BeppoSAX satellite observed $\eta$ Car on June 21--23, 2000
with its Narrow Field Instruments (NFI). NFI include a Low Energy 
Concentrator Spectrometer (LECS) (Parmar et al. 1997), and three 
Medium Energy Concentrator Spectrometers (MECS) (Boella et al. 
1997) at the foci of four X--ray telescopes, 
a Phoswich Detector System (PDS) (Frontera et al. 1997), 
and an High Pressure Gas Scintillation Proportional Counter 
(HPGSPC) (Manzo et al. 1997). 
LECS and MECS units have imaging capabilities and cover
the 0.1--10 keV and 1.5--10 keV energy ranges, respectively. 
PDS covers the 12--200 keV band, while HPGSP the 7--60 keV energy
range. They are both collimated instrument with a field of view of 
1.3 and 1.1 degrees, respectively. 
The PDS collimators are rocked back and forth by about 3$^{\circ}$ 
to allow the simultaneous monitoring of the source and background. 
During 2000 HPGSP was switched off.
$\eta$ Car has been observed for a nominal exposure time of 100 ks. 
The effective exposures were 80.8 ks for 
MECS no. 2 and 3 (MECS no.1 was not in operation),
and 33.4 ks for PDS. The corresponding count rates were, 
MECS: 0.639$\pm$0.003, and PDS: 0.165$\pm$0.032 sec$^{-1}$. 
The LECS image will not be discussed here, 
for the paper is devoted to the analysis of the compact 
hard component which dominates beyond $\sim$2 keV.
The observation date corresponds to phase 1.457 of the 
spectral variation cycle of 5.53 y (Damineli et al. 2000), 
assuming $\Phi = 1$ for the 1998.0 minimum.

Figure 1 shows the MECS images of the region around $\eta$ Car 
using photon events selected from three different energy bands, 
in order to trace the softer and harder X-ray sources in the field, 
and to map the regions emitting in the 6.7 keV iron line.  
It is evident in the figure that 
$\eta$ Car is the hardest and most luminous object in the region. 
Some other sources in the field present a non negligible residual 
in the 7.5--10 keV range and at the iron line, the most interesting 
one being the Wolf--Rayet star HD 93162/WR 25 (WN6+O4), 
to the West of $\eta$ Car.

This observation, in particular, shows that $\eta$ Car is the
main, and, most probably, unique contributor to the flux
observed with the PDS instrument. As discussed in Paper I 
the PDS field of view also includes the X--ray pulsar 
1E 10148.1--5937; but, according to Oosterbroek et al. (1998) 
and Tiengo et al. (2002) its spectrum is softer  
and the flux above 10 keV is much fainter than that of $\eta$ Car. 
Also the lack of intense diffuse emission in the high energy MECS band 
seems not to support the hypothesis of an extended very high energy 
emission due to the interaction of the winds from the massive
stars in the Carina Nebula. 

\section{Spectral analysis} 
The MECS spectrum of $\eta$ Car was extracted within a circular 
region centred on the star with a radius of 4 arcmin. 
To take into account for possible contamination from the 
nebular X--ray emission, we extracted the background 
from a coronal region around $\eta$ Car 
with a 4.3 arcmin inner radius and 11.3 arcmin outer radius.
We also excluded from the selected area two circular areas 
of 4 arcmin in radius centred on the strong 
X--ray sources HD 93162 (WN6$+$O4) and HD 93250 (O3.5V).
The background subtracted 
spectrum of $\eta$ Car and the local nebular spectrum 
normalised to the same area are shown in Fig. 2.  
The MECS background subtracted spectrum of $\eta$ Car 
has been rebinned in order to have at least 20 source 
photons per energy bin.

The background subtracted PDS data have been rebinned up to 50 
keV following the standard procedure. Above 50 keV the data were 
binned in order to have a 2$\sigma$ count rate in each bin. 
As shown in Figure 3, $\eta$ Car has been detected at the 
$\ge$3$\sigma$ level up to 45 keV, and at the 2$\sigma$ level 
at higher energies. 
$\eta$ Car had already been detected by PDS in December 1996 
and December 1999, but only in the 13--20 keV range
due to the shorter exposure times. 
A PDS upper limit was derived from the March 1998 observation.
Table 1 reports the 13--20 keV PDS count rates for the four 
BeppoSAX observations. 

% 
%======================= Figure 2 ============================ 
\begin{figure}
\centering
\includegraphics[width=7.0cm,angle=270]{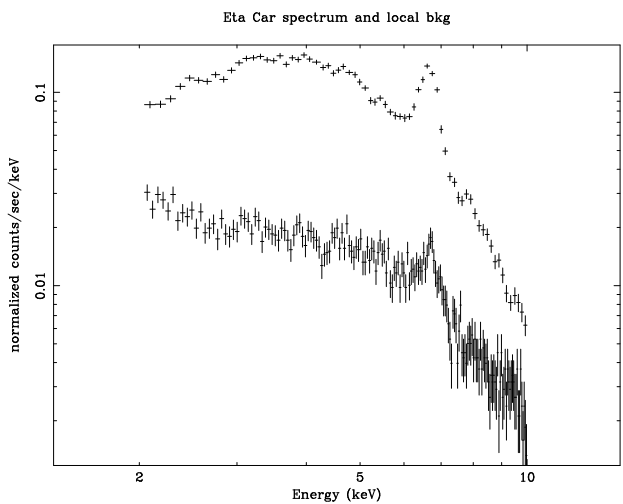} 
\caption{The BeppoSAX MECS background subtracted count rates 
(in s$^{-1}$ keV$^{-1}$) of $\eta$ Car 
in June 2000 compared with the local nebular background spectrum.  
}
\end{figure}
%=============================================================

%======================= Figure 3 ============================ 
\begin{figure}
\centering
\includegraphics[width=7.0cm,angle=270]{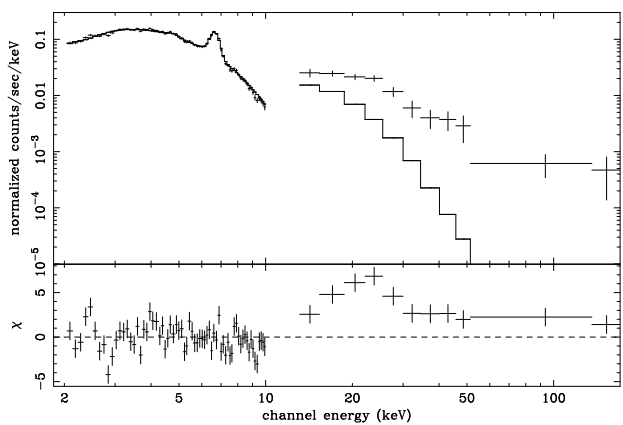} 
\caption{Upper panel: Results of the fit of the June 2000 
BeppoSAX MECS background subtracted spectrum with a one 
temperature (5.51 keV) MEKAL model. 
The spectral model is extrapolated to the PDS energy range in 
order to show the high energy flux excess. 
The residuals are plotted in the lower panel.
}
\end{figure}
%=============================================================
%

The extracted MECS$+$PDS spectrum of June 2000 was analysed 
using the XSPEC spectral fitting software package. 
We started by analysing MECS alone in the 2--10 keV energy 
range which is dominated by the hard core component. 
The MECS spectrum was first fitted with a thermal model (MEKAL) 
leaving iron abundance free to vary. 
We obtained the following best fit values of the parameters:
kT$_h$=5.51$\pm$0.25 keV, 
NH$_h$=4.82$\pm$0.12$\times 10^{22}$ cm$^{-2}$, and 
a logarithmic iron abundance of [Fe/H]=--0.15$\pm$0.02. 
The reduced $\chi^2$ was 1.749 (for 64 degrees of freedom), 
partly due to the small 
instrumental energy mismatch at 6.7 keV discussed in Paper I. 
The source temperature is close to that measured in the MECS 
spectra of December 1996 and March 1998, and reported in Paper I. 
In June 2000 the 2--10 keV energy distribution is 
close to that of December 1996, with nearly the same NH$_h$.
As discussed in Paper I, NH$_h$ was much larger in March 1998 
during $\eta$ Car's egress from the X--ray minimum event 
($\Phi = 1.05$; see also Figure 6 below). 

As described in Paper I, we have also tried to model 
the observed MECS spectrum using an absorbed 
bremsstrahlung model and a Gaussian line.
We obtained for the iron line 
a peak energy of 6.743$\pm$0.001 keV and an equivalent width
of 0.964 keV, with values for kT$_h$ (5.70$\pm$0.14) 
and NH$_h$ ((4.91$\pm$0.10)$\times 10^{22}$ cm$^{-2}$) similar 
to those derived from the previous MEKAL model. 
As reported by Viotti et al. (2002b), 
The equivalent width of the iron line is consistent with  
that of December 1996 (1.05 keV) and of December 1999 (1.01 keV), 
but smaller than that measured in March 1998 (see Table 1). 
The peak energy of the line, corrected for the 
$+$0.04 keV miscalibration discussed in Paper I, 
suggests that the line is mostly emitted by the hot plasma, without 
any important contribution from the 6.4 keV fluorescent line. 
The reduced $\chi^2$ was 1.96 in the range 2--10 keV. 
The integrated absorbed and unabsorbed fluxes in the 2--10 keV 
energy range are: $f_x$=6.45$\times 10^{-11}$ \erg and 
$f_x^o$=9.34$\times 10^{-11}$ \erg, respectively. The latter 
corresponds to 19 \LS\ for $\eta$ Car's distance of 2.6 kpc.

Then, we have analysed MECS and PDS simultaneously, allowing a 
PDS/MECS normalisation factor of 0.85, to take into account the 
miscalibration between the two instruments. 
This correcting factor was also used in Paper I,
though not explicitly indicated. As shown in Fig. 3,  
PDS counts are well above the extrapolation to higher 
energies of the MECS best fit thermal model. 
The difference between observed and expected flux 
increases with energy, suggesting the presence of 
an additional very high energy component. 

In order to account for the high energy tail, we tried 
to fit the 2--150 keV spectrum with two thermal components. 
However, the fit of MECS+PDS with two freely varying temperatures 
always gives unrealistic results. 
Therefore, we fit MECS+PDS with frozen kT(1)=5.51 keV and 
NH(1)=4.82$\times 10^{22}$ cm$^{-2}$, 
and put constraints to kT(2) to be close to 10 keV. 
We thus derived:  kT(2)=14$\pm$5 keV, and  
NH(2)=(543$\pm$22)$\times 10^{22}$ cm$^{-2}$, 
with a reduced $\chi^2$ of 2.33 (Fig. 4).
However, this result is not convincing, not only because 
the fit runs below the observed flux above 30 keV, 
but also because an unlikely very 
high NH is required for the higher temperature component. 
This suggests that the high energy spectrum of $\eta$ Car 
could be non--thermal. 

Therefore, we tried to reproduce the 2--150 keV spectrum 
with a combination of thermal bremsstrahlung (with frozen 
kT=5.70 keV) and power law components, having the same 
absorption column density (4.82$\times 10^{22}$ cm$^{-2}$), 
and a Gaussian line at 6.7 keV. 
The best fit model is shown in Figure 5. 
The non--thermal component has a best fit photon index 
of 2.42$\pm$.04 (reduced $\chi^2$=2.41, for 79 $dof$). 
If produced in a shock, this photon index would imply a 
compression ratio for the shock front of $\chi$=3.1, thus the 
shock should be relatively weak (White \& Chen 1994). 
The nonthermal component in the 13--150 keV range has an 
integrated energy flux of 5.51$\times 10^{-11}$ \erg ~ (11.6 \LS),
a value comparable to that of the thermal 5 keV component.

We have also attempted to fit the overall MECS$+$PDS spectrum with 
a single absorbed power law spectrum plus a Gaussian line. 
Unexpectedly, quite a good fit was obtained (excluding the photon 
events below 2 keV) with a best fit photon index of 2.467$\pm$0.024, 
NH=(6.46$\pm$0.12)$\times$10$^{22}$, an Fe line centred at 
6.733$\pm$0.007 keV, and a reduced $\chi^2$ of 2.404. 
However, this solution appears unrealistic,  
in particular because it would imply that the iron emission 
be mostly the 6.4 keV fluorescent line, in disagreement 
with the measured peak energy of the feature 
(E$_{\rm cor}$=6.70$\pm$0.01 keV). 
Therefore, the thermal plus power law model appears 
more likely to us. 
Incidentally, we notice that the two power law spectra have 
nearly the same slope, independently of the energy range considered. 

%========================== Table 1 =========================
\begin{table*}
\caption[]{ Summary of the 
BeppoSAX observations during 1996--2000. 
Phase one corresponds to the spectroscopic minimum of 1998.0  
}  
\begin{flushleft}
\begin{tabular}{cccccrcc}
\hline
\noalign{\smallskip} 
\noalign{\medskip} 
date  &   JD   &phase &PDS (13-20 keV) &Weq(Fe-k) 
&f$_{\rm X}^{\circ}$(2-10 keV)&  kT$_h$ & NH$_h$\\
      &2400000+&      &    s$^{-1}$    &  keV      
& erg cm$^{-2}$ s$^{-1}$ & keV & cm$^{-2}$  \\
\noalign{\medskip} 
\hline 
\noalign{\medskip} 
1997.00& 50447 & 0.83 & 0.15$\pm$.05 &1.05$\pm$0.06
&9.4 10$^{-11}$&4.8$\pm$0.1& 4.3$\pm$0.1 \\
1998.21& 50891 & 1.05 &-0.04$\pm$.05~&1.41$\pm$0.09
&12.7 10$^{-11}$&4.4$\pm$0.2&15.4$\pm$0.4  \\
2000.00& 51545 & 1.37 & 0.16$\pm$.04 &1.01$\pm$0.06
&8.4 10$^{-11}$&5.1$\pm$0.2& 3.8$\pm$0.1 \\
2000.48& 51718 & 1.46 & 0.17$\pm$.03 &0.96$\pm$0.04 
&9.3 10$^{-11}$&5.5$\pm$0.3& 4.9$\pm$0.1 \\ 
\noalign{\medskip} 
\hline 
\noalign {\medskip} 
\end{tabular}
\end{flushleft} 
\end{table*} 
%===============================================================

\section{Discussion}
The high energy X--ray spectrum of $\eta$ Car is challenging, 
both because such an energetic and powerful 
source has so far never been seen in any other stellar--like source,
and for no current model for non thermal emission seems 
at present applicable to the case of $\eta$ Car. 

In the $\eta$ Car's binary model it is assumed that the thermal 
(5 keV) X--ray emission is originated in the hot shocked gas produced
by collision between the winds of the two stellar components 
(e.g. Pittard et al. 1998, Ishibashi et al. 1999). 
The shocked region should be placed at a distance 
from a few to many AU from the surface of the S Dor star 
during the highly eccentric orbital motion of the system. 
In this model, the plasma temperature is linked to the wind 
velocities, and would be unable to produce directly 
the highest energy photons seen in $\eta$ Car. 

The power law spectrum exhibited by $\eta$ Car above 10 keV 
indicates the presence of an additional non--thermal source. 
In principle, non--thermal emission can be explained by several 
physical processes, most of which are, however, hardly 
compatible with current models of $\eta$ Car. 
A plausible, but still difficult to accept process, would be inverse 
Compton scattering (IC) of low frequency photons by high energy 
electrons. In the case of $\eta$ Car a powerful photon source 
could be represented by the intense 
ultraviolet radiation from the stellar components 
of the binary system, probably from the hot secondary, 
since most likely the UV photons from the S Dor primary do not
emerge from its dense wind. 
Stellar UV photons scattered by relativistic electrons would 
finally emerge with energies in the X-- and $\gamma$--ray 
range, and may carry away much of the energy pumped 
into the electrons at shocks. 

%======================= Figure 4 ============================ 
\begin{figure}
\centering
\includegraphics[width=5.5cm,angle=270]{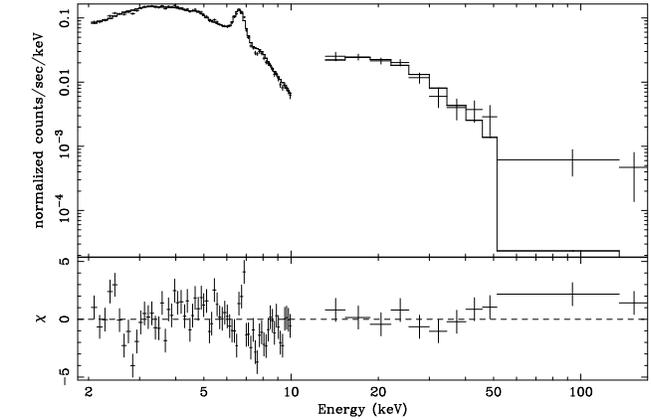} 
\caption{Upper panel: Results of the fit of the BeppoSAX MECS and PDS 
(2--150 keV) spectrum of $\eta$ Car in June 2000  
with a two--temperature (5.5 keV and 14 keV) MEKAL model.
The residuals are plotted in the lower panel.
}
\end{figure}
%=============================================================

%======================= Figure 5 ============================ 
\begin{figure}
\centering
\includegraphics[width=6.0cm,angle=270]{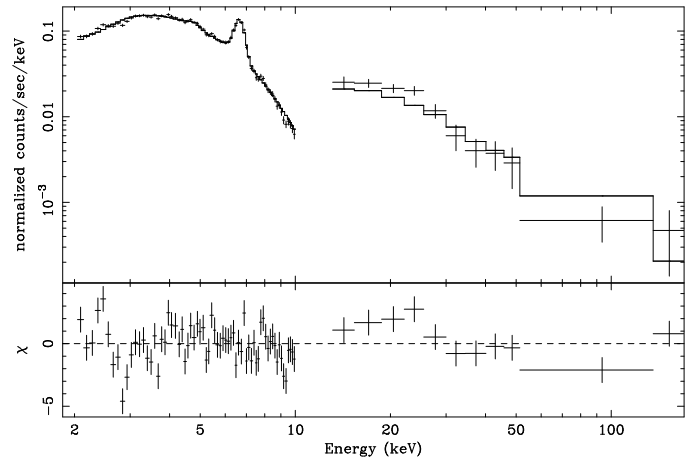} 
\caption{The BeppoSAX MECS and PDS (2--150 keV) spectrum of 
$\eta$ Car fitted with a 5.70 keV thermal bremsstrahlung 
plus a power law spectrum with photon index 2.42, 
and a Gaussian line centred at 6.70 keV.   
} 
\end{figure}
%=============================================================

This process would require the presence of relativistic 
electrons with $\gamma \sim 100$. 
According to White (1985) and Chen \&\ White (1991) 
in the winds of early--type stars electrons 
can be accelerated to relativistic energies 
via the first--order Fermi mechanism in strong isothermal shocks 
generated by radiation driven instabilities in the winds. 
On the other hand, Jardine et al. (1996) 
have shown that in a colliding wind binary 
system, electrons can be accelerated up to relativistic energies 
in a current sheet, formed when the magnetic field carried in 
the stellar winds are forced together as the winds collide.  
In the case of $\eta$ Car there is up to now no direct or indirect 
evidence of the presence of very high energy electrons. 

An indirect evidence could be non--thermal radio emission, 
like that detected in galactic and extragalactic high energy 
X--ray sources. However, non--thermal radio emission centered 
on $\eta$ Car's stellar core, has not been so far detected 
although we cannot exclude its presence. 
In fact, were it present it would lie below the surface that is 
optically thick due to free--free emission. 
Actually, the whole central region of $\eta$ Car is optically 
thick in radio wavelengths to several arcseconds out. 
Since the non--thermal emission should not extend 
that far away from the shock, it should be completely 
hidden by the thermal emission from the dense ionised stellar winds 
(Stephen White, priv. comm.). 
One should also consider that, as recently discussed by Dougherty 
et al. (2003), in colliding wind binary systems the 
non--thermal radio emission could be weakened by self--absorption,  
and inverse Compton cooling. 

Non--thermal radio emission has been detected in a number 
of massive binaries, suggesting the presence of relativistic 
electrons with power--law spectra, which could result in hard 
X-- and $\gamma$--ray non-thermal flux via IC scattering 
(e.g. Benaglia \& Romero 2003).
Among them, the WR$+$O binary system WR 140 bears some resemblances 
to the $\eta$ Car system for its long period (7.9 yr), and 
high orbital eccentricity (e.g. Marchenko et al. 2003), 
and for the presence of a variable IR excess attributed to
dust formation near periastron (e.g. Williams et al. 1987). 
The X--ray emission of WR 140 is one order of magnitude smaller
than that of $\eta$ Car, but, like $\eta$ Car, it suffers a strong  
absorption excess near phases 0.03 and 0.08 (Zhekov \& Skinner 2000). 
If the physical process of production of non--thermal radio 
emission (that is of high energy electrons) in WR 140 is also 
working in $\eta$ Car, we would expect from $\eta$ Car a very 
strong non--thermal radio emission. As discussed above, 
its absence is probably due to the heavy obscuration 
by the thermal absorption from the dense stellar winds.  
Indeed, the crucial point of the spectrum $\eta$ Car's core 
radio emission remains at least at present unsolved. 

Self--Comptonisation is a mechanism proposed to explain 
non--thermal emission in AGNs, as well as in galactic 
compact X--ray sources. 
As for instance discussed by Hua \&\ Titarchuk (1995),
quasi--power law X--ray spectra can be produced by diffusion 
of low frequency photons by optically thick relativistic 
plasma clouds. In this framework,
we might assume that the spectrum of $\eta$ Car above 10 keV 
be produced by scattering of the 5 keV thermal radiation itself 
by high energy ($>$10 keV) electrons present  
in the shocked region. 
This seems to us a more promising process as it does not require 
highly relativistic electrons as in the IC case.  

The study of the spectral variability along the 5.53 yr cycle 
could be a complementary way to tackle the problem of 
the origin of the high energy X--ray tail of $\eta$ Car. 
Table 1 reports the PDS count rates 
and the best fit parameters of the thermal component 
during the four BeppoSAX observations. 
We recall that, according to the RossiXTE observations,
in December 1997 $\eta$ Car underwent a deep X--ray eclipse,
which is currently associated with the periastron passage 
of the suggested binary system (Ishibashi et al. 1999). 
The X--ray eclipse lasted until the end of February 1998. 
Hence, the BeppoSAX observation of mid March 1998 
was made during $\eta$ Car's egress from the eclipse. 

Fig. 6 compares the BeppoSAX MECS spectra observed during the
four epochs. For the sake of homogeneity with the other observations, 
only MECS units 2 and 3 have been 
considered for the December 1996 spectrum. 
It is evident in the figure that beyond the iron feature the four 
spectra nearly overlap, also suggesting that the highly absorbed 
March 1998 spectrum had about the same temperature 
and unabsorbed flux as in the other three epochs. 

As for PDS, it turns out that on three cases the PDS count rate 
in the 13--20 keV range was the same within the data uncertainty. 
As shown in Table 1, in March 1998 $\eta$ Car was not detected 
with PDS. The 3$\sigma$ upper limit of 0.15 s$^{-1}$ is equal 
to the flux measured in the other three observations,
but we think that this coincidence is accidental,
and that in March 1998 the high energy tail was weak. 
Most probably, this component followed the trend of the 
thermal one, and largely weakened during the 
1998.0 X--ray eclipse. 
Our March 1998 observation suggests that, 
while the thermal component had already recovered 
its luminosity, although with a residual large absorption,
the high energy tail was still weak. 
It should have recovered its luminosity at a later time 
(but well before phase 1.37). 
It may be noticed that a long recovering time was also found 
in the visual spectrum for the high excitation emission lines 
(e.g. [Ne \III]), which, as discussed in Paper I, 
were still very weak in May 1998.  
The limited time--coverage does not allow us to conlude 
whether there is a physical link between the two phenomena,
but it would be worthwhile investigating it in the future. 

%======================= Figure 6 ============================ 
\begin{figure}
\centering
\includegraphics[width=9.0cm,angle=0]{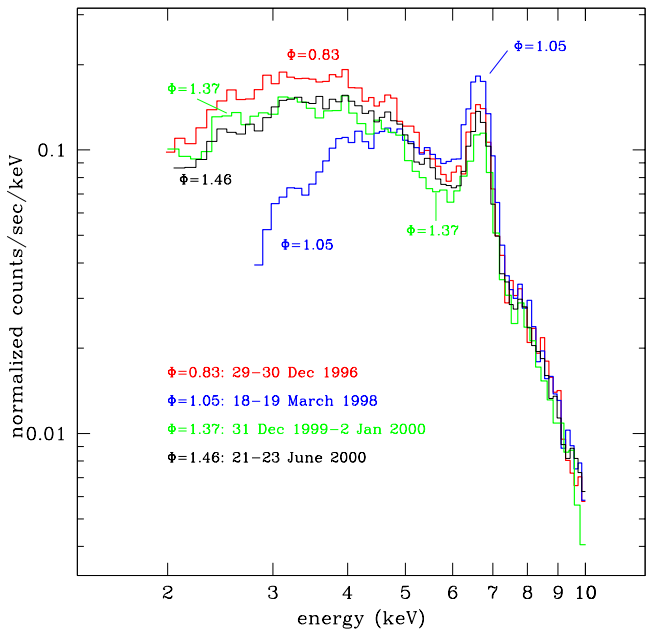} 
\caption{Plot of the BeppoSAX MECS23 spectra of $\eta$ 
Car during December 1996--June 2000. The phases of the 5.53 yr 
cycle are indicated. Note the larger flux of the iron line in March 
1998 ($\Phi = 1.05$) when the spectrum was strongly absorbed. 
At energies $>$7 keV the four spectra basically overlap each other. 
}
\end{figure}
%=============================================================

\section{Conclusion} 
We have presented the first in--depth analysis of the spectrum 
of the $\eta$ Car system above 10 keV, based on a long exposure 
BeppoSAX observation. The power law best fit suggests a 
non--thermal origin of the hard X--ray tail.
The integrated 13--150 keV luminosity ($\sim$12 \LS), is comparable  
to the luminosity of the 2--10 keV thermal component (19 \LS), 
and suggests the presence of a very effective formation process. 

Two models have been considered in the light of the proposed 
binary nature of $\eta$ Car. 
In one model the high energy tail is produced by inverse Compton 
scattering of the UV stellar photons by relativistic 
electrons produced in the wind of the primary star, or 
in the shocked colliding wind region. 
Alternatively, the high energy photons are produced by 
self--Comptonisation of the thermal 5 keV emission from relativistic 
electrons with energies much lower than in the previous case. 
Though, as suggested by the referee, 
the weakness of the high energy tail in March 1998 
could be better explained as high inverse Compton cooling 
of the reletivistc electrons during periastron passage, 
when the colliding winds shock is closer to the stars.  
Neither model has so far enough support from observations, like 
detection of non--thermal radio emission from the central source, a 
point which would deserve new very high resolution radio observations.

A crucial point for understanding the nature of the high energy tail
of $\eta$ Car, would also be to investigate the slope of the spectrum 
at higher energies and to measure the high energy cutoff of the
spectrum, which is related to the energy of the scattering particles.
Our PDS observations allow us to determine a lower limit to the cutoff 
energy of $\sim$50 keV. INTEGRAL observations might allow measurement  
of the X--ray spectrum of $\eta$ Car above 100 keV, and to determine 
up to what energy the power law spectrum is extending.

Finally, new high energy ($>$10 keV) X--ray observations, 
e.g. with the foreseen ASTRO--E satellite,  
of $\eta$ Car near the periastron passage of the system will 
provide a clue of where the non--thermal source is located. 
Our March 1998 upper limit might suggest a recovering  
time slower than that of the thermal source, as observed 
in the optical spectra in the high energy emission lines. 
It would important to investigate the physical link 
beteen the two phenomena, and whether that behaviour is associated 
with the high NH$_h$ value found also at the eclipse egress.

\acknowledgements{ 
We are grateful to the other BeppoSAX team members at ASDC for help 
in acquiring and reducing the data. 
Thanks are due to L. Piro, V. F. Polcaro and Stephen White 
for discussions, and 
to an anonymous referee for constructive suggestions. 
This work is partly based on contract I/R/053/02 of the Italian 
Space Agency (ASI).}


\begin{thebibliography}{}
\bibitem{}
 Andriesse, C. D., Donn, B. D., Viotti, R. 1978, MNRAS, 185, 771 
\bibitem{}
 Benaglia, P., Romero, G. E. 2003, A\&A, 399, 1121 
\bibitem{}
 Boella, G., Chiappetti, L., Conti, G., et al. 1997, A\&AS, 122, 327 
\bibitem{} 
 Chen, W., White, R. L. 1991, ApJ, 381, L63 
\bibitem{}
  Chen, W., White, R. L. 1994, Astrophys. Space Sci 221, 259 
\bibitem{}
 Corcoran, M. F., Ishibashi, K., Swank, J. H., Petre, R. 2001,
 ApJ, 547, 1034 
\bibitem{}
  Damineli, A., Kaufer, A., Stahl, O., Lopes, D. F., de Araujo, F. X.
 2000, ApJ, 528, L101 
\bibitem{}
 Dougherty, S. M., Pittard, J. M., Kasian, L., et al. 
 2003, A \& A, 409, 217 
\bibitem{}
Duncan, R. A., White, S. M. 2003, MNRAS, 338, 425 
\bibitem{}
 Frontera, F., Costa, E., Dal Fiume, D., et al.  1997, A\&AS, 122, 357
\bibitem{}
  Hillier, D. J., Davidson, K., Ishibashi, K., Gull, T. 2001, 
 ApJ, 553, 837
\bibitem{}
  Hua, X.-M., Titarchuk, L., 1995, ApJ, 449, 188 
\bibitem{} 
  Ishibashi, K., Corcoran, M. F., Davidson, K., et al. 
1999, ApJ, 524, 983 
\bibitem{}
  Jardine, M., Allen, H. R., Pollock, A. M. T. 1996, A\&A, 314, 594 
\bibitem{}
Manzo, G., Giarrusso, S., Santangelo, A., et al. 1997, A\&AS, 
  122, 341 
\bibitem{}
 Marchenko, S. V., Moffat, A. F. J., Ballereau, D., et al. 2003, 
ApJ, 596, 1295 
\bibitem{} 
 Oosterbroek, T., Parmar, A. N., Mereghetti, S., Israel, G. L. 
 1998, A\&A, 334, 925 
\bibitem{}
 Parmar, A., Martin, D. D. E., Bavdaz, M., et al. 1997, A\&AS, 122, 309 
\bibitem{}
 Pittard, J. M., Stevens, I. R., Corcoran, M. F., 
 Ishibashi, K., 1998 MNRAS, 299, L5 
\bibitem{} 
 Pittard, J. M., Corcoran, M. F. 2002, A\&A, 383, 636 
\bibitem{} 
  Rebecchi, S., Viotti, R. F., Grandi, P., Antonelli, L. A., 
 Corcoran, M. F., Damineli, A., Rossi, C. 2001, in 
 X--Ray Astronomy 2000, ed. R. Giacconi, S. Serio, \& L. Stella, 
 ASP Conf. Ser. Vol. 234, 107 
\bibitem{}
  Tiengo, A., G\"ohler, E., Staubert, R., Mereghetti, S. 2002,
 A\&A, 383, 182 
\bibitem{} 
van Boekel, R., Kervella, P., Sch\"oller, M., et al. 2003, 
  A\&A, 410, L37
\bibitem{} 
Viotti, R. 1995, Rev. Mex. Astr. Ap. Conf. Series Vol. 2, 1 
\bibitem{} 
  Viotti, R., Corcoran, M. F., Damineli, A., Grandi, P. 1998, 
 Proc. Symp. The Active X-Ray Sky, L. Scarsi et al. eds., 
 Nuclear Phys. B (Proc. Suppl.), 69/1-3, 36 
\bibitem{}
  Viotti, R. F., Antonelli, L. A., Corcoran, M. F., 
 Damineli, A., Grandi, P., Muller, J. M., Rebecchi, S., 
Rossi, C., Villada, M. 2002a, A\&A, 385. 874 (Paper I) 
\bibitem{}
  Viotti, R. F., Antonelli, L. A., Rebecchi, S., Rossi, C. 
 2002b, Journal of Astrophysics and Astronomy, Vol.23, pp.19-22 
\bibitem{} 
  White, R.L. 1985, ApJ, 289, 698 
\bibitem{}
  White, R.L., Chen, W. 1994, Astrophys. Space Sci., 221, 295
\bibitem{} 
  Williams, P. M., van der Hicht, K. A., van der Woerd, H., 
 et al. 1987, in Instibilities in Luminous Early type stars, ed.
% H.J.G.L.M 
H. J. G. et al. Lamers, \& C. W. H. de Loore (Dordrecht: Reidel), 221 
\bibitem{} 
  Zhekov, S. A., Skinner, S. L. 2000, ApJ, 538, 808 
\end{thebibliography}
\end{document}